\documentclass[journal,comsoc]{IEEEtran}
\usepackage[T1]{fontenc}% optional T1 font encoding

\ifCLASSINFOpdf
\else
\fi
\usepackage{amsmath}
\usepackage[cmintegrals]{newtxmath}
\usepackage{cite}
\usepackage{algorithmic}
\usepackage{graphicx}
\usepackage{textcomp}
\usepackage{xcolor}
\usepackage{array,multirow}
\def\BibTeX{{\rm B\kern-.05em{\sc i\kern-.025em b}\kern-.08em
    T\kern-.1667em\lower.7ex\hbox{E}\kern-.125emX}}

\begin{document}
\title{HELPS for Emergency Location Service:
Hyper-Enhanced Local Positioning System
}

\author{Hichan~Moon,
        Hyosoon~Park,
        and~Jiwon~Seo% <-this % stops a space
%\thanks{H. Moon is with the Department of Electronic Engineering, Hanyang University, Seoul, 04763 Republic of Korea e-mail: hcmoon@hanyang.ac.kr.}% <-this % stops a space
%\thanks{H. Park is with Infoseize Systems Co., Seoul, Korea.}% <-this % stops a space
%\thanks{J. Seo is with the School of Integrated Technology, Yonsei University, Incheon, 21983 Republic of Korea e-mail:.}% <-this % stops a space
%\thanks{Manuscript received September 00, 2021;}
}


\markboth{IEEE Wireless Communications,~Vol.~00, No.~0, April~2024}%
{Moon \MakeLowercase{\textit{et al.}}: HELPS for Emergency Location Service}



\maketitle

\begin{abstract}
In this study, we propose a novel positioning and searching system for emergency location services, namely the hyper-enhanced local positioning system (HELPS), which is applicable to all mobile phone users, including legacy feature phone users. In the case of an emergency, rescuers are dispatched with portable signal measurement equipment around the estimated location of the emergency caller. Each signal measurement device measures the uplink signal from the mobile phone of the caller. After calculating the rough location of the caller’s mobile phone based on these measurements, rescuers can efficiently search for the caller using the received uplink signal strength. Thus, the positioning accuracy in a conventional sense is not a limitation for rescuers in finding the caller. HELPS is not a traditional positioning system but rather a system with humans in the loop designed to reduce search time in emergencies. HELPS can provide emergency location information even in environments where the GPS or Wi-Fi is not functional. Furthermore, for HELPS operation, no hardware changes or software installations are required on the caller's mobile phone.
\end{abstract}

%\begin{IEEEkeywords}
%wireless positioning, emergency location service, positioning and searching system, E911.
%\end{IEEEkeywords}

% \fi
\IEEEpeerreviewmaketitle

\section*{Introduction}
\IEEEPARstart{T}{he} location information of an emergency caller is crucial to dispatch rescuers effectively, but there are numerous technical challenges in locating a wireless caller with sufficient accuracy \cite{Corral-De-Witt18:From}. 
Locating a wireless emergency caller has been actively studied since the U.S. Federal Communications Commission (FCC) issued a two-phase approach for wireless E911 through the \textit{Report and Order and Further Notice of Proposed Rulemaking} (CC Docket No. 94-102) in 1996. 
The horizontal location accuracy requirement for handset-based solutions specified in the FCC’s \textit{Third Report and Order} was 50 m for 67\% of calls and 150 m for 95\% of calls by 2001. 

This requirement motivated manufacturers to include a global positioning system (GPS) and other global navigation satellite system (GNSS) chipsets within mobile phones. Although GNSS provides a high positioning accuracy in open-sky environments, its accuracy can significantly degrade in urban multipath environments \cite{Lee23:Nonlinear}. Furthermore, it is almost impossible to acquire and track GNSS signals indoors because of the very weak signal strength \cite{Kassas17:I}. 

Considering that a large volume of wireless emergency calls are made indoors, the FCC’s \textit{Fourth Report and Order} in 2015 adopted the horizontal accuracy requirement of 50 m for 80\% of calls by 2021, including indoor calls. 
There are various indoor positioning technologies available, which include assisted-GNSS (A-GNSS), pattern matching (also called fingerprinting) of opportunistic radio frequency (RF) signals such as wireless fidelity (Wi-Fi), Bluetooth, and long-term evolution (LTE), and additional infrastructure-based approaches \cite{Zafari19:A_Survey, Tian20:RF}. 
Nevertheless, determining the caller's vertical location, which is crucial for rapid location identification in multi-story buildings, remains particularly challenging.

According to the FCC’s most recent \textit{Sixth Report and Order on Reconsideration} in 2020, a vertical accuracy of $\pm 3$ m for 80\% of calls made from z-axis capable devices is required in each of the top 25 and 50 cellular market areas (CMAs) by 2021 and 2023, respectively, and nationwide by 2025. The \textit{Stage Za Test Bed Report} of the 9-1-1 Location Technologies Test Bed, LLC, in 2020 mentioned, ``no commercial Z-axis solutions have yet been validated to meet the requirements'' although some encouraging test results were reported. 

The 50 m horizontal and $\pm 3$ m vertical accuracies, once provided, greatly reduce emergency response times. However, it should be noted that location accuracy is an important \textit{starting point} for rescuers to find the caller. Even when the required location accuracy is provided for a certain wireless call, it is still challenging to locate the caller within a tall building in a dense urban area. There can be many buildings within 50 m, several floors within $\pm 3$ m, and many rooms on the floor. Thus, it is challenging for rescuers to \textit{search} for and open the caller’s room in minimal time. 

Therefore, we turn our attention from increasing the positioning accuracy in a traditional sense to reducing the overall \textit{search time} because it is the ultimate goal of E911. 
In this vein, we propose a novel technology designed to reduce the search time even when the initial location uncertainty is high. 
Importantly, this technology is applicable to all mobile phone users, including those with legacy feature phones.

\section*{Related Work}

There have been numerous studies regarding indoor and outdoor localization methods \cite{Zafari19:A_Survey, Tian20:RF, Zhu20:Indoor, Chukhno22:D2D-Based, Albanese21:First, Ko21:V2X}. 
The primary purpose of these studies was to increase positioning accuracy. 
Certain technologies, such as A-GNSS and Wi-Fi fingerprinting \cite{Zafari19:A_Survey, Zhu20:Indoor}, have already been implemented within commercial smartphones, significantly improving the accuracy of smartphone positioning, especially in indoor environments where GNSS signals are unavailable. 
Other technologies, such as D2D-based cooperative localization and localization in 5G NR densification scenarios \cite{Chukhno22:D2D-Based, Kanhere21:Position}, have shown the potential to further enhance smartphone positioning accuracy. 
However, these technologies have not yet been implemented in commercial smartphones.

Despite these research efforts, first responders or rescuers today cannot guarantee finding an emergency caller who is unable to verbally communicate their position in an urban area within the ``golden hour.'' 
This is because the performance of these positioning methods is highly dependent on infrastructure installation, signal environment around the caller, and the specific model (generation, capability, and installed software) of the caller's mobile phone. 
In some cases, the caller's mobile phone may not be a smartphone, although most recent positioning technologies assume smartphone users.

In reality, first responders today spend a substantial amount of time on the last-mile search for an emergency caller after receiving the caller's position information from wireless carriers. 
According to FCC Commissioner Mignon Clyburn, ``it is estimated that as many as 10,000 lives could be saved each year, if the 911 emergency dispatching system were able to reach callers just one minute faster.'' 

It is very challenging to pinpoint the correct room where the caller is located in a tall building in an urban area. 
Unless a ``positioning'' system can consistently provide sufficient horizontal and vertical positioning accuracy to locate an indoor caller, which is an ambitious goal given current technologies, a ``searching'' system will be beneficial in reducing the search time for rescuers. 
It should be noted that such a searching system does not compete with a positioning system; instead, they complement each other. 
After determining the caller's location through positioning, the searching process begins. 
Thus, higher positioning accuracy will alleviate the burden of searching since the search space will be smaller.

A familiar example of a searching system is the ``find my phone'' feature of a smartphone. 
If this feature were used for emergency rescue, when a rescuer arrives at the indicated location, they must activate the sound alarm on the target smartphone and rely on audible sound to locate the phone. 
However, in this case, a hostile individual near the emergency caller may discover the smartphone before the rescuer and damage it. 
Even when a hostile person is not present, the sound alarm may not be audible outside closed doors of apartments or houses.
Therefore, the ``find my phone'' feature of current smartphones is not an appropriate searching system for emergency rescue.

The searching system that we propose in this article utilizes the uplink signals transmitted from a mobile phone, as they are much ``quieter and stronger'' than the sound alarm of a smartphone. 
A rescuer equipped with a specially designed receiver can find the caller much more successfully and efficiently than a rescuer relying solely on their ears and the ``find my phone'' feature in practice. 
Since the proposed system leverages the basic call connection and protocol between a target mobile phone and a base station, it is applicable even to legacy feature phones without any hardware and software modifications. 
To the best of the authors' knowledge, the system presented in this article represents the first uplink cellular signal-based emergency search system designed for all types of mobile phones.

\section*{Proposed System}

\begin{figure}
  \centering
  \includegraphics[width=1.0\linewidth]{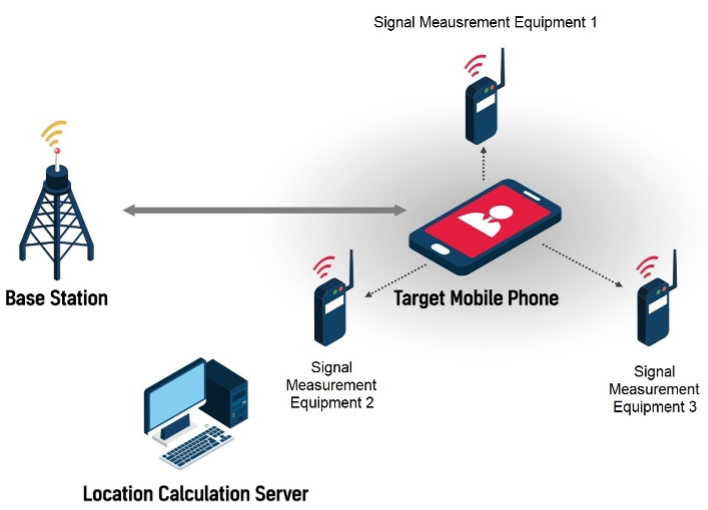}
  \caption{Basic operation of HELPS. Each SME measures uplink signals from a target mobile phone, and LCS generates the location information of the target mobile phone based on the measurements.}
  \label{fig:BasicOperation}
\end{figure}

To reduce the search time given a certain initial position uncertainty following the application of existing positioning methods, we propose a hyper-enhanced local positioning system (HELPS) for emergency location services.
Figure \ref{fig:BasicOperation} illustrates the overall operation of HELPS. 
A base station connects a call with a target mobile phone belonging to an emergency caller. 
The base station sends the uplink channel configuration information to a portable signal measurement equipment (SME) that measures signals from the target mobile phone. 
After measuring the uplink cellular signals transmitted from the target mobile phone, each SME sends the measurement reports to a location calculation server (LCS). 
The LCS collects measurement reports from SMEs and generates the location information of the target mobile phone. 

One of the key innovations of HELPS is the portable SMEs. 
There has been a previous positioning method, namely the uplink time difference of arrival (UTDOA), based on the measurement of uplink signals transmitted from a mobile phone \cite{DelPeral-Rosado18:Survey}.
However, because it relies on stationary infrastructures or measurement devices, positioning accuracy degrades depending on the location of the mobile phone.
Furthermore, because UTDOA requires time synchronization between signal measurement devices, signal measurement devices should utilize GNSS or other devices for time synchronization.
This requirement limits the possible locations of signal measurement devices for UTDOA, thus impeding the enhancement of positioning accuracy and caller search operations through the dynamic deployment of these devices. 
Due to these limitations, UTDOA has not been deployed commercially.

In contrast, with HELPS, each rescuer dispatched to find an emergency caller uses a portable device to measure the uplink cellular signal from the caller's mobile phone \cite{Moon:Position}. 
For time synchronization with a cellular network, each SME includes a downlink receiver of a cellular system. 
Therefore, it is possible to deploy SMEs at any locations within a cellular network without concerns about time synchronization.

\subsection*{Components of HELPS}
HELPS consists of four main components: a target mobile phone, signal measurement equipment, a location calculation server, and a cellular network to support HELPS operation.

\subsubsection*{Target Mobile Phone}
A target mobile phone is the mobile device of an emergency caller, which is the search target. 
It can be any mobile phone capable of connecting to a cellular network. 
HELPS is based on the basic functionality and protocol of a cellular phone.
Therefore, no hardware modifications or software installations are required on a target mobile phone.

\subsubsection*{Signal Measurement Equipment (SME)}
An SME is portable equipment designed to measure the cellular signals transmitted from the target mobile phone. 
The concept of HELPS is applicable to any cellular signals, but our SME prototype utilizes LTE signals. 
The SME prototype consists of a signal measurement unit (SMU) and a smartphone.

The SMU comprises a downlink receiver, an uplink receiver,
and a controller. 
Since the SMU does not include a transmitter in cellular bands, it does not generate any interference or degrade the performance of the cellular system. 
An SMU acquires the channel configuration information of the target mobile phone’s uplink signal and measures the uplink signal transmitted from the target mobile phone.
The SMU is connected to a rescuer’s smartphone using a universal serial bus (USB) or Wi-Fi. 
The connected smartphone is used to exchange the measurement and location information with the LCS.

\begin{figure}
  \centering
  \includegraphics[width=0.9\linewidth]{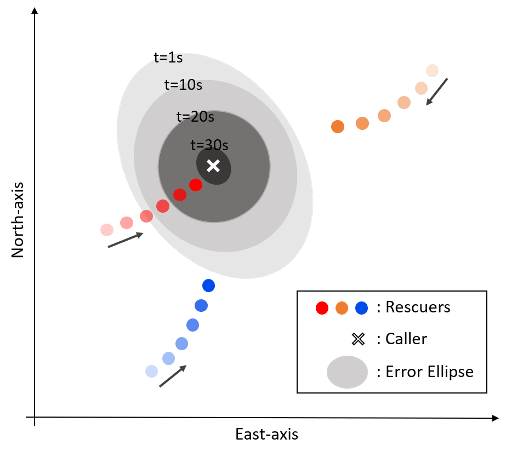}
  \caption{Conceptual illustration of positioning accuracy improvement. As rescuers approach the emergency caller and estimate the caller's position based on continuous measurements of the uplink signals, the location estimation error ellipse gradually reduces in size, resulting in higher positioning accuracy.}
  \label{fig:AccImproveConcept}
\end{figure}

\subsubsection*{Location Calculation Server (LCS)}
The LCS collects the measurement results for the target mobile phone sent by SMEs. 
It generates the location information of the target mobile phone based on the received measurement results and transmits this location information to the SMEs. 
To calculate the location information of the target mobile phone, the LCS can utilize the received power and propagation delay measured at each SME. 
The LCS is capable of generating a contour map of received power levels for searching purposes and calculating the target's location based on the measured propagation delay for positioning purposes.

\subsubsection*{Cellular Network}
A base station establishes a call connection with the target mobile phone and configures a periodic uplink signal. 
For instance, in the case of an LTE system, an evolved Node-B (eNB) orders a target mobile phone to transmit a physical uplink shared channel (PUSCH) or physical random access channel (PRACH) periodically. 
The channel configuration information for the uplink signal of the target mobile phone is transmitted to the SMEs through the LCS.

\begin{figure}
  \centering
  \includegraphics[width=1.0\linewidth]{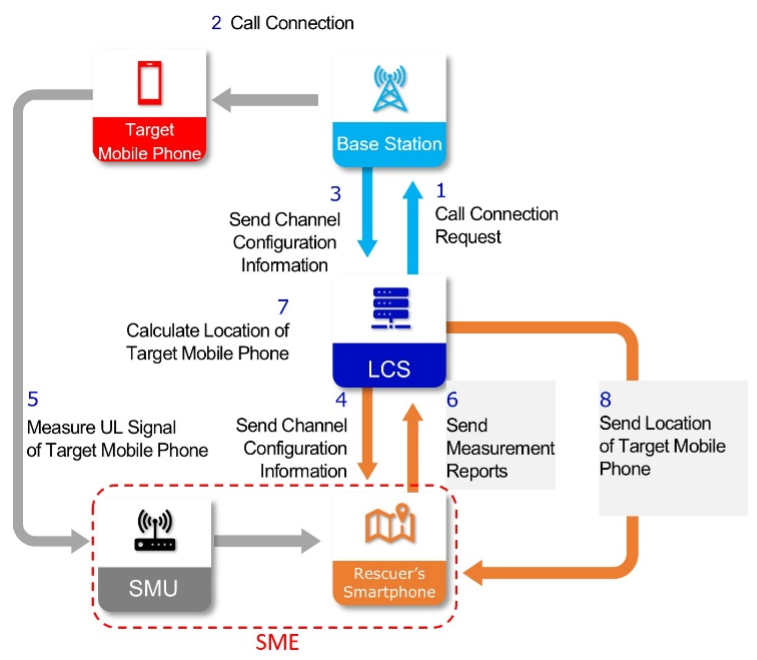}
  \caption{Operational scenario of HELPS.}
  \label{fig:OperationalScenario}
\end{figure}

\begin{figure*}
  \centering
  \includegraphics[width=0.9\linewidth]{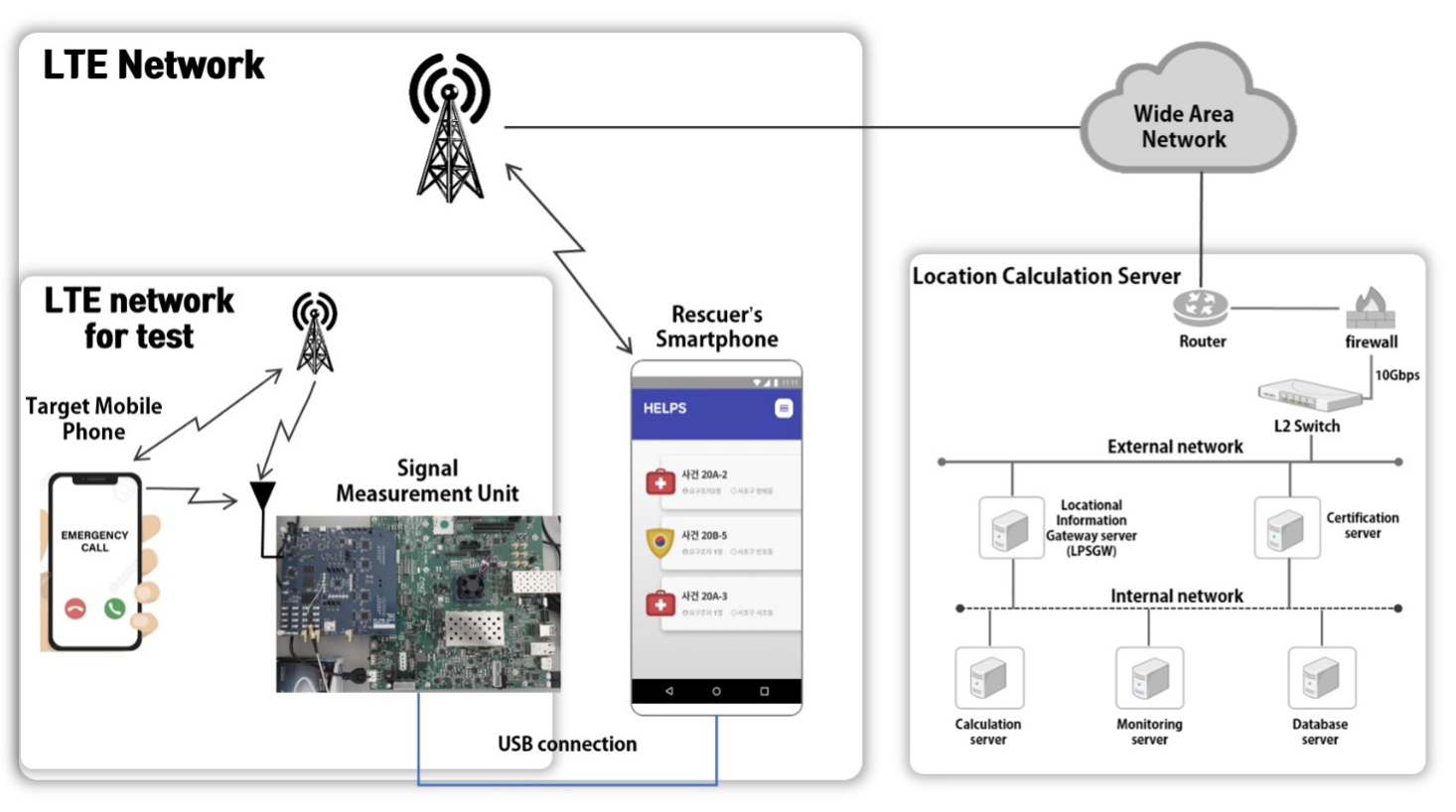}
  \caption{Architecture of HELPS prototype system.}
  \label{fig:HELPS_Architecture}
\end{figure*}

\subsection*{Operational Scenario of HELPS}

When an emergency call is made from the target mobile phone of an emergency caller, initial location information can be obtained using existing positioning systems. 
These existing positioning methods \cite{Zafari19:A_Survey} include cellular tower-based methods and GPS/Wi-Fi-based methods on the target mobile phone. 
After obtaining the initial location information of the target mobile phone, a team of rescuers equipped with portable SMEs is dispatched to the vicinity of the location. 
Upon arriving at a location close to the target mobile phone, a call connection is established between the target mobile phone and a nearby base station. 
The HELPS call connection request can be initiated from either an SME or an LCS. 
Subsequently, a HELPS positioning and searching session commences for the target mobile phone.
It should be noted that there is no connection required between the target mobile phone and SMEs. 
SMEs are passive detection and measurement devices.

Leveraging the mobility of SMEs, it is possible to achieve a more accurate position estimation of the target mobile phone as the SMEs approach the target, as illustrated in Fig. \ref{fig:AccImproveConcept}.
However, there are no limitations on the positioning methods applied to obtain the position estimate.
The position estimate can be obtained through the HELPS positioning session or other available methods such as cellular tower-based methods.
The uncertainty in position estimation, represented by the error covariance of a position estimation filter, defines a search boundary for the upcoming emergency caller search, as explained in the following subsection.
The improved positioning accuracy obtained as the SMEs approach the target can reduce the search boundary and consequently the overall search time.

Figure \ref{fig:OperationalScenario} illustrates the operational scenario of HELPS. 
A call connection request is generated from the LCS. 
If a HELPS call connection request is received, a base station establishes a call connection with the target mobile phone. 
For example, a base station can command periodic uplink transmission from a target mobile phone. 
The information of the uplink channel configuration is sent to the LCS. 
(If the base station allocated predefined uplink resources to a target mobile phone, it is possible to skip this process of sending the channel configuration information.)
Subsequently, the LCS forwards the uplink channel configuration information to SMEs that are attempting to locate the target mobile phone. 
Each SME measures the uplink signal from the target mobile phone based on the channel configuration information. 
After measuring the uplink signal transmitted from the target mobile phone, each SME reports the measurement results to the LCS. 
The LCS generates the location information of the target mobile phone based on the measurement reports and sends the estimated location to SMEs. 
Each SME displays the estimated location of the target mobile phone on a map. 
Now, the HELPS search session is ready to begin.

\subsection*{Caller Search Session of HELPS} 

Once the search boundary is defined according to the location estimation uncertainty, rescuers initiate the HELPS search session using handheld SMEs capable of measuring the received signal strength indicator (RSSI) values of the LTE uplink signals from the caller. 
This search process is analogous to finding metal using a ``metal detector'' because the signal strength from the target mobile phone increases as the SME approaches the caller. 
Even with a brute force search, the caller will eventually be found, but more efficient search algorithms are desired to reduce the search time. 
Considering urban environments, we propose a two-phase approach, consisting of a building search phase and a floor and room search phase.

\begin{figure*}
  \centering
  \includegraphics[width=0.6\linewidth]{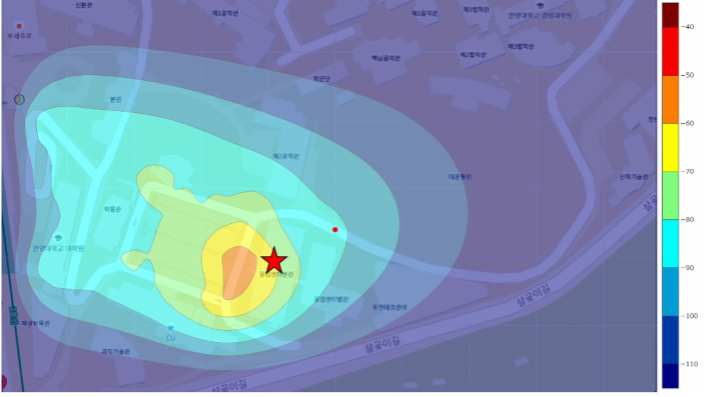}
  \caption{Example of building search result.}
  \label{fig:BldgSearchResult}
\end{figure*}

\subsubsection*{Building Search}
A preliminary location estimate for an emergency caller is currently provided by wireless carriers. 
A more accurate initial location estimate would reduce the search boundary and, consequently, decrease the search time. 
However, a preliminary estimate from a cellular tower-based method \cite{Schloemann16:Toward} is also acceptable for defining a search boundary.

The LCS defines the search boundary based on the initial location estimate, divides the search area within the boundary according to the number of dispatchable rescuers, and assigns designated areas and routes to each rescuer. 
Each rescuer drives along the given routes, and the SME continuously measures the RSSI using an omnidirectional antenna, sending this information to the LCS. 
The LCS creates an RSSI contour map over the search area by interpolating the RSSI measurements from the SMEs.

Once the RSSI contour map is created, the LCS sends the location with the highest RSSI value in the map to the rescuers' SMEs. 
Subsequently, the rescuers move to the designated location and switch the SMEs to the directional search mode. 
In the directional search mode, the SME uses a directional antenna and displays the measured RSSI value in a specified direction. 
While multiple buildings may be in the vicinity of the highest-RSSI location, the manual directional search utilizing a directional antenna can precisely identify the building from which the caller is transmitting LTE uplink signals.

\subsubsection*{Floor and Room Search}
The concept of searching the caller's floor and room within a designated building is similar to the building search scenario. 
The LCS divides the floors of the building according to the number of rescuers and assigns specific floors to each rescuer. 
Subsequently, rescuers walk through the corridors, while SMEs equipped with omnidirectional antennas continuously measure RSSIs and send this information to the LCS. 
The LCS then interpolates the RSSI values and generates an RSSI contour map for each floor. 
Based on the RSSI maps of all floors, the LCS determines the location with the highest RSSI, which is then sent to the SMEs. 
Upon reaching the location with the highest RSSI, rescuers switch their SMEs to the directional search mode to precisely identify the room where the caller is located.

\section*{Prototype}

The architecture of the HELPS prototype system implemented for the proof of concept is illustrated in Fig. \ref{fig:HELPS_Architecture}. 
Each SME consists of an SMU implemented with a field-programmable gate array (FPGA) and a commercial smartphone. 
An application program is installed on the rescuer’s smartphone to facilitate the operation of HELPS. 
For the proof of concept, an LTE network was constructed at a test site. 
A HELPS call connection was established between a target mobile phone and the LTE eNB using the test LTE network. 
SMEs communicate with LCS using a commercial LTE network. 
LCS comprises external and internal networks. 
The external network includes a location information gateway server and a certification server, while the internal network includes a calculation server, a monitoring server, and a database server. 

The screen of a rescuer’s smartphone displays the created RSSI contour map along with the estimated position of a target mobile phone, with the uncertainty of estimation represented by a circle around the estimated position.
Additionally, the display shows the locations of other colleague rescuers who are also searching for the same target mobile phone. 
This enables each rescuer to access location information not only for the target mobile phone but also for their fellow rescuers. 
With this location information and the RSSI contour map, the rescuers can proceed towards the target mobile phone. 

To validate the concept and performance of HELPS, a test LTE system was deployed at Hanyang University, Seoul, Korea. 
This LTE system consists of two commercial radio units (RUs) located in two separate buildings and covers the entire campus and its surrounding areas. 
A commercial digital unit (DU) controls the RUs. 
The LTE system uses a 10 MHz bandwidth in both the downlink and uplink. 
The center frequencies of the downlink and uplink signals are 793 and 738 MHz, respectively, within LTE Band 28. 
For the target mobile phone of an emergency caller, a commercial LTE mobile phone was utilized.

In the context of HELPS operation, the transmission power and the period of uplink signal transmission play crucial roles in determining the power consumption of the target mobile phone. 
After considering the trade-off between performance and power consumption, a base station can adjust these parameters.

For the results presented in this paper, the target mobile phone transmitted a 1 ms PUSCH with approximately 7.5 MHz bandwidth. 
With this setting, on average, the transmission power of PUSCH was measured to be approximately 10 dB higher than that of a standard voice call. 
If the period of uplink signal transmission is set to about 80 ms, the power consumption of a target mobile phone during HELPS operation remains roughly equivalent to that of a typical voice call. 
An SME measured the uplink signal sent from the target mobile phone using two demodulation reference signal (DM-RS) symbols in a 1 ms PUSCH subframe with two antennas.

\section*{Experimental Results}

The building search method of HELPS was validated using the prototype system. 
An area of approximately 250 m $\times$ 300 m was considered as the search area, implying that the initial position estimation from wireless carriers had very high uncertainty. 
Figure \ref{fig:BldgSearchResult} shows an example of the building search results. 
The LCS continuously updated the RSSI contour map based on the RSSI measurements from the rescuers’ SMEs. 
As depicted in Fig. \ref{fig:BldgSearchResult}, the peak location of the contours was close to the actual location of the target mobile phone, indicated by a red star. 
Given the presence of multiple buildings around the peak contour, a rescuer performed a manual directional search using a directional antenna to identify the correct building.

Using the HELPS testbed, we conducted 50 building search experiments by randomly placing the target mobile phone within one of 25 buildings, each consisting of five or more floors. 
The RSSI contour map was generated based on the RSSI measurements from three SMEs in three vehicles. 
A building search was considered successful when the building with the target mobile phone was located without entering the building. 
With more than an 80\% probability, the building with the target mobile phone was located within three minutes.

It is worth noting that this result is based on the assumption that the initial horizontal position error was approximately 125 m. 
If a certain positioning system can achieve a horizontal position error better than 50 m, as required by the E911 mandate, it would significantly reduce the search time needed to pinpoint the correct building.
In this scenario, the new search area would be only 16\% of the previous search area ($50^2 / 125^2 = 0.16$).
 
\begin{figure}
  \centering
  \includegraphics[width=0.9\linewidth]{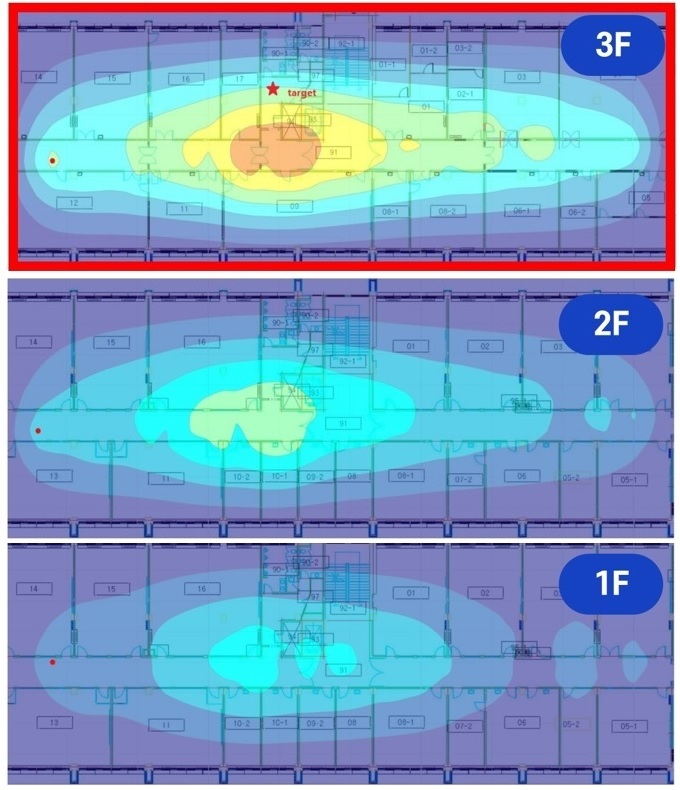}
  \caption{Example of floor and room search result.}
  \label{fig:RoomSearchResult}
\end{figure}

After locating the correct building, the subsequent task is to identify the specific floor and room containing the target mobile phone. 
Figure \ref{fig:RoomSearchResult} illustrates the RSSI contour maps generated by the LCS when a mobile phone is located on the third floor of a five-story building.
In the figure, only the RSSI maps for the first, second, and third floors are displayed, as the RSSI maps for the fourth and fifth floors closely resemble those of the second and first floors, respectively.

Rescuers measured the RSSIs of the uplink signals from the target mobile phone using SMEs equipped with omnidirectional antennas. 
The actual location of the target mobile phone is denoted by a red star, which in this case is on the third floor. 
As expected, the peak RSSI contour closely aligned with the location of the target mobile phone. 
Upon reaching the peak RSSI area, the rescuer changed the SME to directional search mode to pinpoint the correct room where the target mobile phone was located.

With the HELPS testbed, we conducted 100 floor and room search experiments by randomly placing the target mobile phone within one of approximately 100 rooms in a building. 
Each room's dimensions are approximately 4 m $\times$ 8 m. 
A room search was considered successful when the room containing the target mobile phone was located without the need to open its door.
It was possible to locate the room with the target mobile phone within three minutes with an approximately 90\% probability. 
It should be noted that positioning accuracy is not a proper metric to evaluate the performance of HELPS. 
HELPS does not intend to find the 3D coordinate of the caller but rather assists the rescuers in quickly locating the caller's room in a dense urban environment.

With state-of-the-art smartphone technology utilizing Wi-Fi and cellular downlink measurement information at a target mobile phone, it took an average of 461 s (812 s for 90\% CDF) to knock on the doors of the rooms within a building, check if the caller was inside, and actually find the caller.
This experiment was performed after obtaining the agreements of the people in the rooms of the office building. 
Therefore, the result was under a very friendly environment. 
However, in real situations, it is possible that people inside the rooms may not answer the knock, and the search time would be significantly longer. 
With HELPS, it took 148 s on average (175 s for 90\% CDF) to find the room where the caller was located without knocking on any door of the rooms.

In our experiments, we assumed that vertical position information was unavailable. 
Consequently, the rescuers searched all the floors. 
If a certain positioning system can provide a vertical accuracy of $\pm 3$ m, as required by the E911 mandate, the search time would be significantly reduced because the rescuers would only need to search approximately three floors, regardless of the number of floors in the building.

\section*{Summary of Innovations} 

The innovations of the proposed HELPS system are summarized as follows. 
\begin{itemize}
	\item \textbf{Paradigm shift from positioning accuracy to search time:} While the E911 requirements specify horizontal and vertical positioning accuracies, the primary objective of E911 is to reduce emergency response time, specifically the search time for rescuers to find the emergency caller. In the case of HELPS, where a caller's mobile phone continuously transmits specific uplink signals to a base station during emergencies, rescuers equipped with handheld SMEs can rapidly find the caller based on received uplink signal strengths within a search boundary. Unlike other positioning technologies, our goal is not to obtain precise geodetic coordinates of the caller but rather to efficiently search for and find the caller within the shortest possible time. This paradigm shift from positioning accuracy to search time has facilitated the development of the HELPS architecture.
	
	\item \textbf{Not limited by accuracy:} In conventional emergency location services, the primary bottleneck arises from the accuracy of the target mobile phone's location. A last-mile search (e.g., knocking on all doors within the area of location uncertainty) can be time-consuming. In contrast, HELPS caller search operation is not limited by location accuracy. Rescuers equipped with portable SMEs can approach the target mobile phone based on uplink signal strength measurements, even when location uncertainty is high.
 
	\item \textbf{Emergency location service to all mobile phones without any modification:} HELPS relies on the basic call connection and protocols between a target mobile phone and a base station without utilizing any smartphone sensors or features. 
    Therefore, no hardware modifications or software installations are necessary for mobile phones. This means that emergency location services can be provided even to low-cost mobile phones. 
    HELPS requires new equipment (i.e., SME) for rescuers to measure the cellular signal transmitted by a target mobile phone and LCS to process these measurements. To enable HELPS operation, software upgrades on the cellular network side are necessary; however, no additional hardware components are required within the cellular networks.
\end{itemize}

\section*{Conclusion}

In this article, we propose HELPS, a novel positioning and searching system designed for emergency location services. 
HELPS offers a highly practical solution to expedite the search for an emergency caller within densely populated urban areas. 
HELPS can provide emergency location services to any mobile phone if it can make a call connection with a neighboring base station, which is a trivial requirement.
Because rescuers can quickly search for an emergency caller based on the measurements of uplink signals from the target mobile phone, the initial positioning accuracy is no longer a bottleneck for emergency location services. 
We have implemented a proof-of-concept system and plan to test a fully operational system within a commercial LTE network in the near future. 
This innovation holds significant promise for improving emergency response times and enhancing the safety of individuals in critical situations.

\section*{Acknowledgment}
The authors would like to express their gratitude to the Ministry of Science and ICT, Korea, for their support in frequency allocation for the LTE test site at Hanyang University. 
Special acknowledgment is extended to Network Business, Samsung Electronics Co., Korea, for providing base station equipment and software to support HELPS operation, and to Korea Telecom Co., Korea, for supplying core network equipment and infrastructure for the HELPS test site.
The authors also recognize the dedicated researchers and engineers who contributed to the design and implementation of the HELPS test system. 
This work was supported by the Institute of Information \& Communications Technology Planning \& Evaluation (IITP) grant funded by the Korean government (KNPA) (2019-0-01291, LTE-based accurate positioning technique for emergency rescue). 
Hichan Moon and Jiwon Seo are the corresponding authors of this article.

\ifCLASSOPTIONcaptionsoff
  \newpage
\fi



\bibliographystyle{./IEEEtran}
\bibliography{./IEEEabrv,./mybibfile}


%\end{IEEEbiography}

\begin{IEEEbiographynophoto}{Hichan Moon}
(hcmoon@hanyang.ac.kr) is a professor in the Department of Electronic Engineering, Hanyang University, Seoul, Korea. He received his B.S. and M.S. degrees in electronics engineering from Seoul National University (Summa Cumme Laude), Seoul, Korea and his Ph.D. degree in electrical engineering in 2004 from Stanford University, Stanford, CA, USA. He was also with Samsung Electronics Co., Korea. In Samsung, he designed mobile station modems for cdma2000, W-CDMA, and LTE. He also worked for standardization of cellular systems. He is the project leader and the main contributor to HELPS.
Since 2021, he has been the chair of Ad Hoc Committee on Mission Critical Communications, IEEE VTS.  
\end{IEEEbiographynophoto}

%\newpage

\begin{IEEEbiographynophoto}{Hyosoon Park} (hyospark@dgu.ac.kr) is currently working as a professor at Dongguk University. He is a co-founder of Infoseize Systems Co., Seoul, Korea.  He received a Ph.D. degree in electrical and electronic engineering, Yonsei University, Korea, in 2005. He was a vice president in the Network Business Division, R\&D team, Samsung Electronics Ltd., Korea. In Samsung, he worked for the development of base station equipment for cdma2000, W-CDMA, WIBRO, and LTE. From 2017 to 2019, he was a professor at the School of Electronic Engineering, Hanyang University, Seoul, Korea.
\end{IEEEbiographynophoto}

\begin{IEEEbiographynophoto}{Jiwon Seo}
(jiwon.seo@yonsei.ac.kr) is a professor with the School of Integrated Technology, Yonsei University, Incheon, Korea, and an adjunct professor with the Department of Convergence IT Engineering, Pohang University of Science and Technology (POSTECH), Pohang, Korea. 
He received his B.S. degree in mechanical engineering (division of aerospace engineering) in 2002 from the Korea Advanced Institute of Science and Technology (KAIST), Daejeon, Korea, and his M.S. degree in aeronautics and astronautics in 2004, an M.S. degree in electrical engineering in 2008, and a Ph.D. degree in aeronautics and astronautics in 2010 from Stanford University, Stanford, CA, USA. He is developing the positioning and searching algorithms of HELPS.
\end{IEEEbiographynophoto}


\vfill


\end{document}